# Bifurcation of Subsonic Gas Flows in the Vicinity of Localized Heat Release Regions

Sergey T. Surzhikov
Russian Academy of Sciences
The Institute for Problems in Mechanics Russian Academy of Sciences
101-1 prospekt Vernadskogo, 119526, Moscow, Russia
surg@ipmnet.ru http://www.ipmnet.ru

*Abstract:* – Time-accurate simulation of gas dynamic structures resulting in viscous, heat conducting and radiating gas flows through localized heat release areas is performed. Cases of intensive heat release in to air at atmospheric pressure are investigated, whereby the gas is heated up to temperatures 3000÷20000 K, so that local differences of density reach tens and hundreds times.

Bifurcation of the gas dynamic structure has been found at certain relation between gas velocity and heat release power.

## 1 Introduction

The necessity for studying gas dynamic structure of subsonic gas flows through localized heat release areas arises in connection with research of laser-supported waves (LSW) [1, 2]. Certain kinds of energy devices are based on the specified phenomenon: the laser plasma generators (LPG) [3] and the laser supported rocket engines (LSRE) [4]. The interest to problems of gas flow interactions with localized heated areas has increased noticeably in physical gas dynamics recently, first of all in connection with the probable applications in practical aerodynamics. The review of corresponding works is presented in [5].

This study concentrates basically on physical regularities of subsonic gas motion through localized heat release areas. The physical statement of the problem was formulated in [6], where not only experimental study of this phenomenon is performed, but also an elementary one-dimensional model is considered. During 1969–1984 years an accumulation of experimental data and development of various one-dimensional models (see review in [1]) took place. The first quasi-two-dimensional model was developed in 1984 [7]. Detailed numerical studies of radiating and gas dynamic processes in LSW and LPG were performed then for consistently more sophisticated models [1, 3, 8–14]. The self-consistent models of gas dynamics and laser heating as a rule were investigated in the above papers. Further progress in the experimental studies is reflected in [1, 15–17]. And, in spite of the fact that the gas dynamic structure has been described in detail, the basic attention was given to research of modes of the LSW existence and speeds of ones moving. Only in last years [13, 14, 19] a necessity for study of the own unsteady structures of the LSW was formulated.

In parallel with study of the LSW the numerical simulation studies in fundamental gas dynamics and astrophysics were carried out [18, 20–30] under the description of gas dynamic phenomena accompanying local heat release in flows. The linear and weakly linear models for generation of gas flows disturbances by means of laser radiation are considered in [21, 22]. Acoustic disturbances in supersonic flows around a laser beam were studied in [18]. A low-level heat release capacity approximation was used for description of two-dimensional steady-state gas flows in [23, 24]. A nonlinear one-dimensional unsteady theory of a thermal layer inside transonic flows of compressed gas is developed in [25]. Use linear and weakly nonlinear approximations allowed to obtain a number of analytical solutions. However, cases of significant release heat power and of pulsed heat releases require use of the numerical methods.

Two-dimensional numerical simulation of unsteady pulse heat release in essentially subsonic gas flow is made in Ref.26. The problem was solved in two stages: first structures of shock waves generated by a pulse heat release were studied; then the slow subsonic flows were simulated.

A numerical study of a strong two-dimensional disturbances generated in a supersonic flow is performed in [27].

A numerical research of gas dynamic processes involving interaction of supersonic flows with continuous and pulsed-periodical cylindrically symmetric heat release source is performed in [28]. It should be stressed that in all above papers unsteady character of gas flows were caused by the pulsed character of the heat release.

Similar investigations were performed also with reference to astrophysical problems. Study interaction of supersonic star's wind flows with X-ray radiation source is performed in [29, 30]. Equations of perfect gas dynamics were used in this case, but the energy release capacity was described by physically proved model of interaction of X-ray radiation with the gas. This study showed that a head shock wave is formed at high-power energy release, and there is an area of lowered density behind the energy source.

A series of papers should also be noted on numerical simulation of effects accompanying a local heat release in the vicinity of streamlined bodies [5, 31].

The problem considered in the following is characterized by a number of distinctive features:

1. The study of unsteady subsonic motion of viscous heat-conducting and radiating gas in area of heat release is based on the full system of the Navier – Stokes equations. Thus, specified non-steady-state modes can be caused not only by the external reasons, but also by own properties of gas flows through the heat release area.

2. Temperature in the heat release areas can achieve $\sim 5000 \div 20000$ K, that is the gas becomes completely dissociated or ionized. Differences in gas densities in the region under consideration may reach $\sim 200$ times, because the pressure in this area a slightly differs from the atmospheric one.

3. The radiative heat transfer, real thermo-physical and transport properties are also taken into account.

A heat release area is considered in the present study as being fixed in space that is any changes of gas dynamic parameters do not influence on the heat release capacity. In reality when the gas is heated by the LSW, situation a slightly different: the heat release capacity depends rather strongly on distribution of gas dynamic parameters [2]. However, in the latter case it is difficult to specify the reason for occurrence of unsteady movements: whether on account of its internal properties, whether on account of periodic change of the heat release configuration. In other words, the fixed heat release area allows exclude influence of variability in the heat release area and, thus, to study regularity of occurrence of the unsteadiness caused by own properties of the gas flow.

## 2 Statement of the problem

The problem schematic is shown in Fig.1.

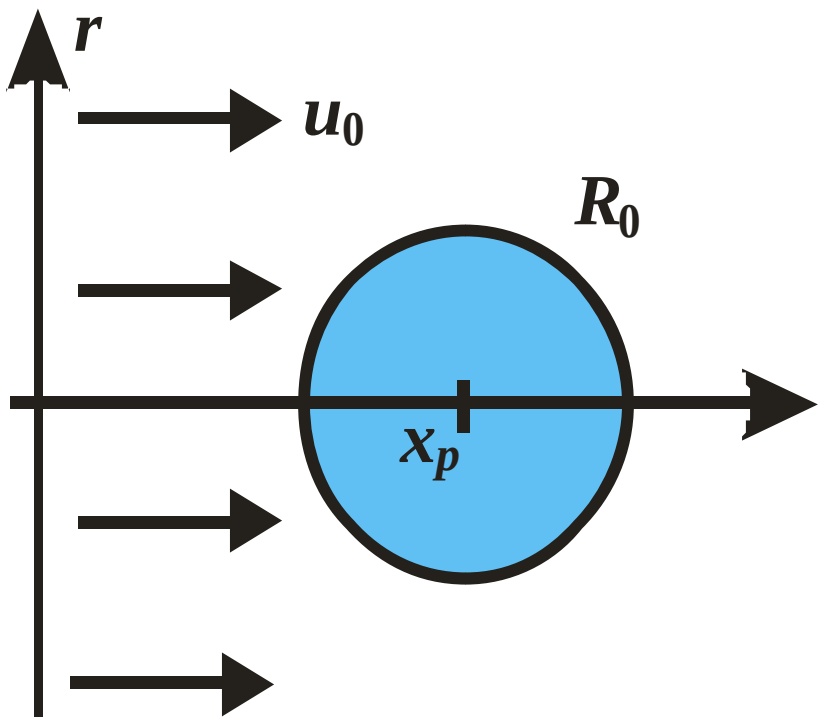

**Fig.1.** Schematic of the problem. The heat release region has a spherical shape. Typical velocities are $u_0 = 10 \div 200$ m/s; typical temperatures inside the heat release region are $T = 1000 \div 20000$ K

On the symmetry axis ($r = 0$) in a point $x_p$ there is a hot area with the given distribution of heat release capacity

$$Q_V = \frac{3Q_0}{4\pi R_0^3} \exp\left[ -\left(\frac{r}{R_0}\right)^4 - \left(\frac{x - x_p}{R_0}\right)^4 \right], \qquad (1)$$

where $R_0$ is the radius of the heat release region, $x_p$ is the axial co-ordinate of its center, $Q_0$ is the heat release power. Note that such form practically coincides with heat release distributions received in self-consistent model of LSW [2]. Again, it should be stressed that the distribution of heat release capacity (Eq.(1)) is determined only by spatial variables and does not depend neither on time, nor from gas dynamic processes.

Parameters of undisturbed gas were used at the entrance of the area under consideration: velocity $u_0$ and temperature $T_0$.

For numerical simulation of subsonic gas flows through heat release regions the following system of the Navier – Stocks equations, mass and energy conservation equations, and also equation of radiative heat transfer in the form of the $P_1$-multi-group approximation is used

$$\frac{\partial \rho}{\partial t} + \mathrm{div}(\rho \mathbf{V}) = 0 \;, \qquad (2)$$

$$\frac{\partial \rho u}{\partial t} + \mathrm{div}(\rho u \mathbf{V}) = -\frac{\partial p}{\partial x} - \frac{2}{3}\frac{\partial}{\partial x}(\mu\, \mathrm{div}\mathbf{V}) + \frac{1}{r}\frac{\partial}{\partial r}\left[ r\mu \left( \frac{\partial u}{\partial r} + \frac{\partial v}{\partial x} \right) \right] + 2\frac{\partial}{\partial x}\left( \mu \frac{\partial u}{\partial x} \right), \qquad (3)$$

$$\frac{\partial \rho v}{\partial t} + \mathrm{div}(\rho v \mathbf{V}) = -\frac{\partial p}{\partial r} - \frac{2}{3}\frac{\partial}{\partial r}(\mu\, \mathrm{div}\mathbf{V}) + \frac{\partial}{\partial x}\left[ \mu \left( \frac{\partial u}{\partial r} + \frac{\partial v}{\partial x} \right) \right] + 2\frac{\partial}{\partial r}\left( \mu \frac{\partial v}{\partial r} \right) + \mu \frac{\partial}{\partial r}\left( \frac{v}{r} \right), \qquad (4)$$

$$\rho c_p \frac{\partial T}{\partial t} + \rho c_p \mathbf{V}\mathrm{grad}T = \mathrm{div}\left(\lambda\, \mathrm{grad}T\right) - Q_{HR} + Q_V , \quad (5)$$

$$Q_{HR} = \sum_{g=1}^{N_g} k_g \left(U_{b,g} - U_g\right) \Delta\omega_g , \quad (6)$$

$$\mathrm{div}\left(3k_g^{-1} \mathrm{grad}\, U_g\right) = -k_g \left(U_{b,g} - U_g\right), \quad (7)$$

$$g = 1, 2, \ldots, N_g ,$$

where $x, r$ are the radial and axial coordinates; $\rho, c_p, T$ are the density, specific heat capacity at constant pressure and temperature; $u, v$ are the axial and radial components of the flow velocity **V**; $p$ is the pressure; $\mu$, $\lambda$ are the coefficients of viscosity and thermal conductivity; $Q_{HR}$ is the volume capacity due to radiation heat transfer; $k, U, U_b$ are the absorption coefficient, radiation volume density of the medium and absolutely block body. Subscript $g$ indicates group properties as obtained by averaging the appropriate spectral characteristics within each of $N_g$ spectral ranges of wave numbers $\Delta\omega_g$.

Validity of the local thermodynamic equilibrium (LTE) is assumed. The gas composition (air in this case) is considered at chemical equilibrium in each point of the calculation area at given temperature and pressure. Because of small speeds of the gas the energy conservation equation does not contain the term representing heat release due to gas compressibility. Temperature dependence of the thermo-physical ($\rho, c_p$), transport ($\mu, \lambda$) and optical ($k_g$) properties of air are used only at atmospheric pressure, as their changes dependence on pressure is insignificant.

At the initial time instant a gaussian temperature distribution with maximal temperature 2000 K is set. The following boundary conditions are used:

at $x = 0$: $V = (u = u_0,\ v = 0)$, $\quad T = T_0$;

at $x = L\ (x \to \infty)$: $\partial f / \partial x = 0$ or $\partial^2 f / \partial x^2 = 0$;

at $r = 0$: $\partial f / \partial r = 0$;

at $r = R_c\ (r \to \infty)$: $f = f_0$ or $\partial f / \partial r = 0$;

where $f = \{u, v, T, U_g\}$.

The values $x_p, L, R_c$ were chosen in numerical experiments from the conditions of weak influence of the boundaries site on the calculation results in the vicinity to the heat release area.

We need not formulate boundary conditions for pressure, because one is excluded from the consideration. The method of the Unsteady Dynamical Variables was applied to the solution of the problem [2, 11]. Thermo-physical and group optical properties of low temperature air plasma were calculated using MONSTER computing system [31].

# 3 Qualitative analysis of the phenomenon

Essentially, the problem is as follows: if at the fixed undisturbed gas velocity, for example at $u_0 = 100$ m/sec, we gradually increase heat release power $Q_V$, each time using just obtained solution for thermo-gasdynamic functions, then at some value $Q_V^*$ a vortical (steady-state or unsteady) gas motion may arise behind the heat release area. If then at some value $Q_V > Q_V^*$ we start to reduce $Q_V$ gradually, then one can find that the vortical gas dynamic structure is preserved at $Q_V < Q_V^*$ down to certain value $Q_V^{**} < Q_V^*$, at which the gas flow once again becomes laminar. Another words, we can say about such well known phenomenon as hysteresis. But, from the point of view of computational fluid dynamics one can say about very significant fact that there is certain range $Q_V^{**} < Q_V < Q_V^*$, where two qualitatively different gas dynamic structures correspond to the same given data ($u_0, Q_V$, and other invariable entrance data).

The specified gas dynamic structures with vortical movement can be observed in wide range of velocities $u_0 \sim 30 \div 200$ m/sec.

One more fact may be of interest too. One of the two obtained solutions, namely one with recoverable vortical motion, is steadier. The numerical experiments have shown, that if the laminar solution is chosen as the initial one, and some indignation is introduced into the flow (for example, change of $Q_V$ by a value $\Delta Q_V$ with in the limits $[Q_V^{**}, Q_V^*]$), then the following two results may be obtained:

At small values $\Delta Q_V$ the solution does not leave the initial branch of the solutions, that is the flow remains laminar;

At large enough $\Delta Q_V$ the solution always converges to the alternative configuration, that is becomes vortical.

If the vortical solution is chosen as the initial one, then indignations of gas dynamic functions do not result in change of the gas dynamic pattern.

It is also necessary to take into account three additional features of the obtained results:

At speeds $u_0 \le 20$ m/sec the bifurcation of gas dynamic structure was not revealed. However it should be stressed, that it is impossible to interpret this conclusion as an absolute, because the calculations were carried out at fixed entrance parameters (pressure, geometry, size of the heat release region, etc.).

Near to border of the conditional stability, that is at $Q_V \approx Q_V^*$ (boundary of the transition «laminar flow – vortical flow») the unsteady solutions are as a rule observed.

With increasing $\Delta Q_V$ reduction of the range $[Q_V^{**}, Q_V^*]$ in computing experiments is observed.

In conclusion of this Section it should be stressed that some examples of bifurcation of gas dynamic structures at low Mach numbers recently were considered in [18, 33].

## 4 Quantitative results of the numerical study

Calculations were carried out for the following entrance speeds $u_0 = 10 \div 200$ m/sec, and for fixed radial and axial coordinates of the heat release region: $R_0 = 0.4$ cm, $x_p = 3$ cm. A maximal value of the heat release power $Q_V$ was varied in the range $2 \div 20$ kW/cm$^3$. The greatest value $Q_V$ was limited by maximal temperature inside the heated area ($T \sim 20000$ K).

Firstly we will consider numerical simulation results obtained at step by step increasing the entrance velocity $u_0$. The computed results obtained for conditions corresponding to stability boundaries of the gas flows will then be considered in detail.

Temperature and axial velocity distributions at $u_0 = 10$ m/sec and $Q_V = 14.6$ kW/cm$^3$ are shown in Fig. 2.

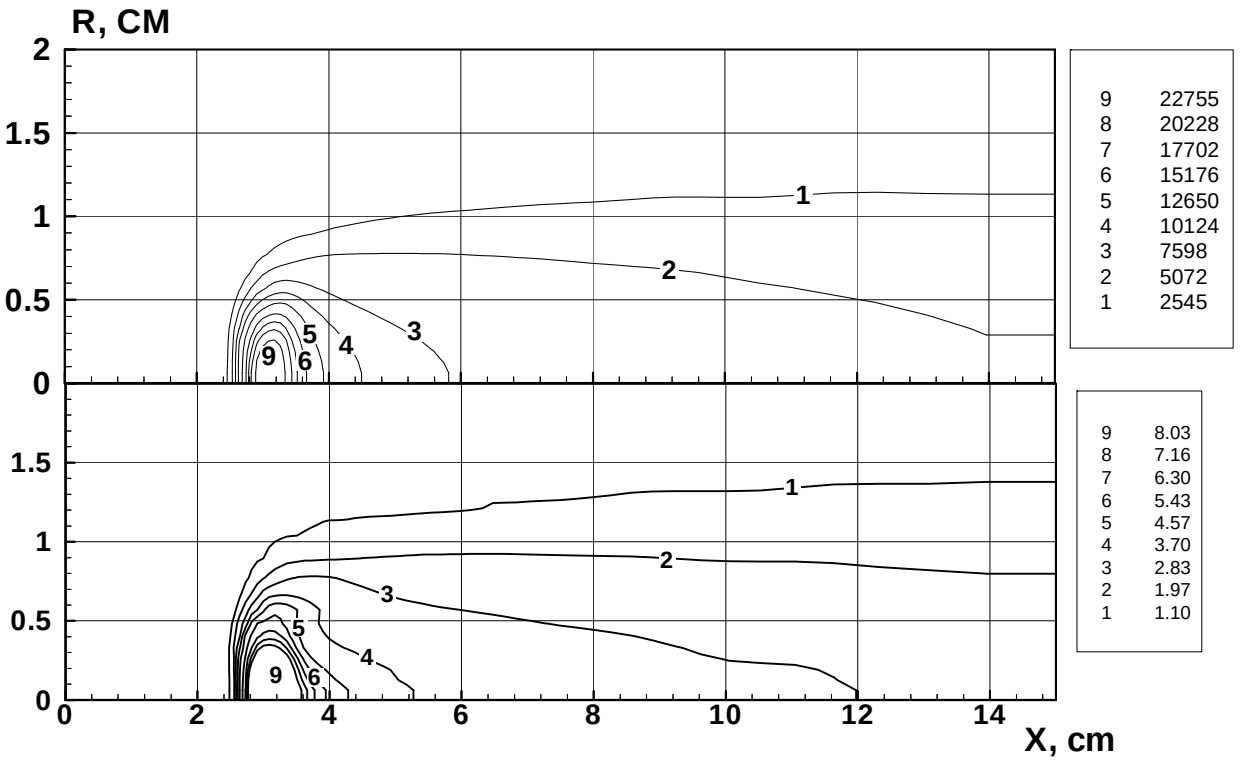

**Fig.2.** Temperature and axial velocity distributions at $u_0 = 10$ m/s and $Q_V = 14.6$ kW/cm$^3$. Set of calculations with **increasing** heat release

The corresponding axial velocity distribution along the symmetry axes is shown in Fig. 3. As it was mentioned above, transition from laminar to vortical mode of gas dynamic structure is not detected in this case. A maximal temperature inside the heated region achieves approximately 24000 K. It should be emphasized that the numerical solution in this case is not completely steady state, and small periodic fluctuations of the velocity and temperature fields are observed. The steady state solution was obtained at essentially smaller heat release power $Q_V \sim 2 \div 5$ kW/cm$^3$.

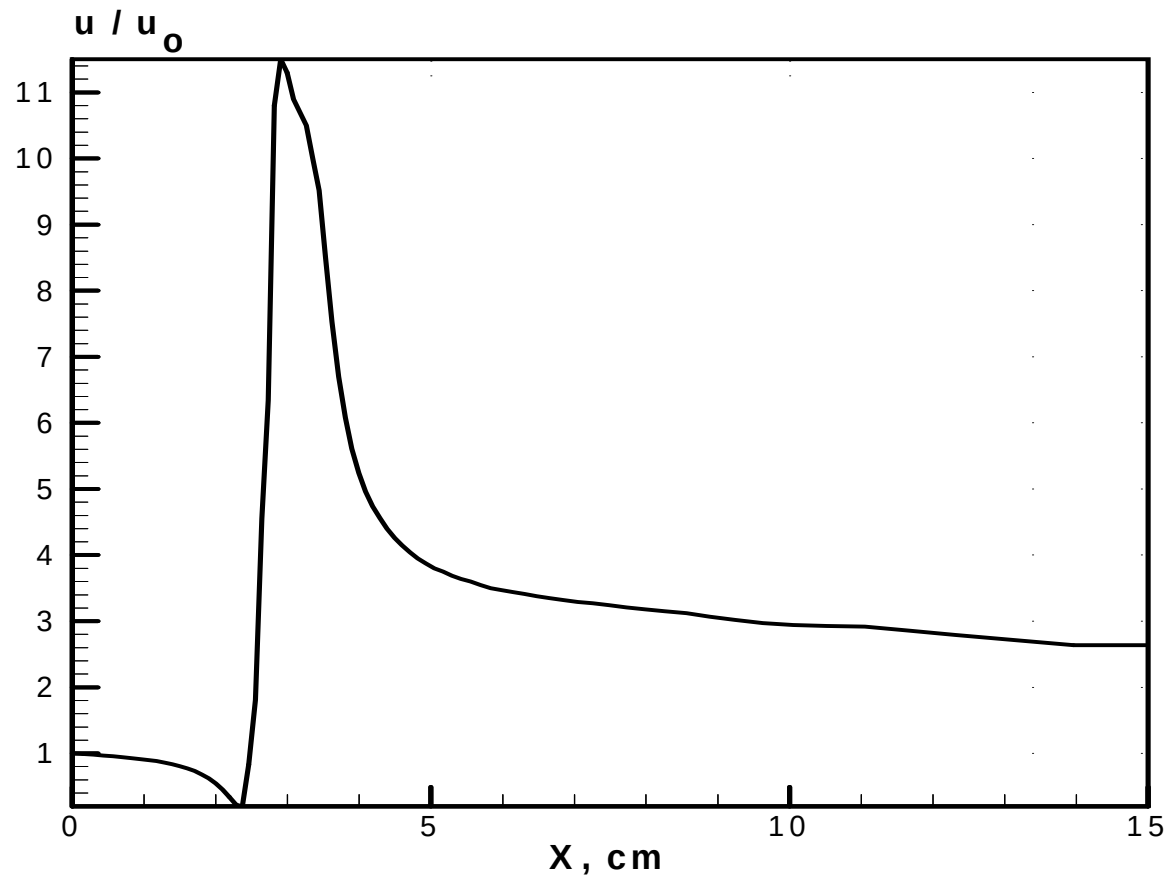

**Fig.3.** Axial velocity distributions at $r = 0$, $u_0 = 10$ m/s and $Q_V = 14.6$ kW/cm$^3$. Set of calculations with **increasing** heat release

Beyond entrance velocity $u_0 = 30$ m/sec the phenomena of bifurcation in the flow was detected, but solutions obtained at this velocity were not stable. Detailed discussion of these results will be presented slightly later. First, we will consider calculated data obtained at velocities $u_0 = 40 \div 200$ m/sec, because these data are stable enough. Temperature and velocity fields for the case of stationary laminar ($Q_V = 2.5$ kW/cm$^3$) and quasi-steady-state vortical flow ($Q_V = 3.0$ kW/cm$^3$) at $u_0 = 40$ m/sec are shown in Fig.4 and 5 respectively. The name «quasi-steady-state» is used here to emphasize that the solution obtained is not perfectly steady-state, but rather contains moderate velocity oscillations.

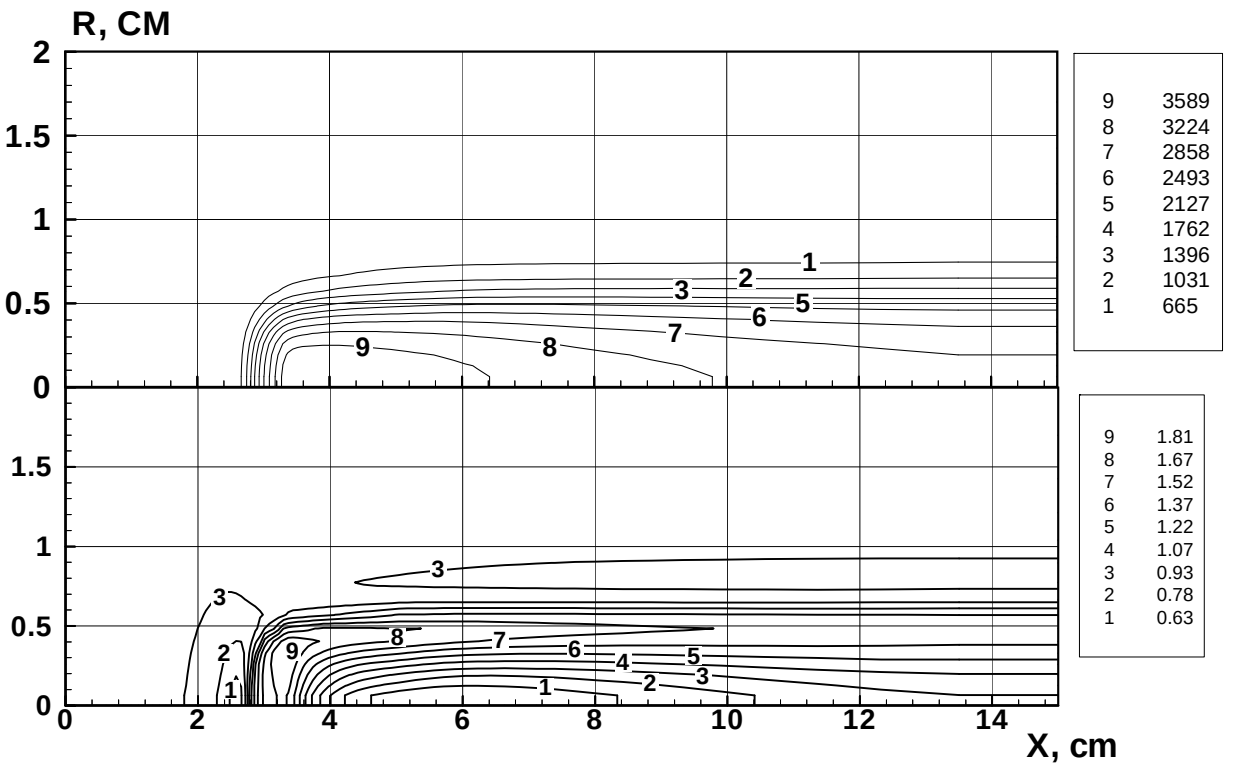

**Fig.4.** Temperature and axial velocity distributions at $u_0 = 40$ m/s and $Q_V = 2.5$ kW/cm$^3$. Set of calculations with **increasing** heat release

The specified solutions were obtained in the set of calculations where $Q_V$ was increased with step $\Delta Q_V = +0.5$ kW/cm$^3$. Maximal temperature inside the heated area achieves 5200 K.

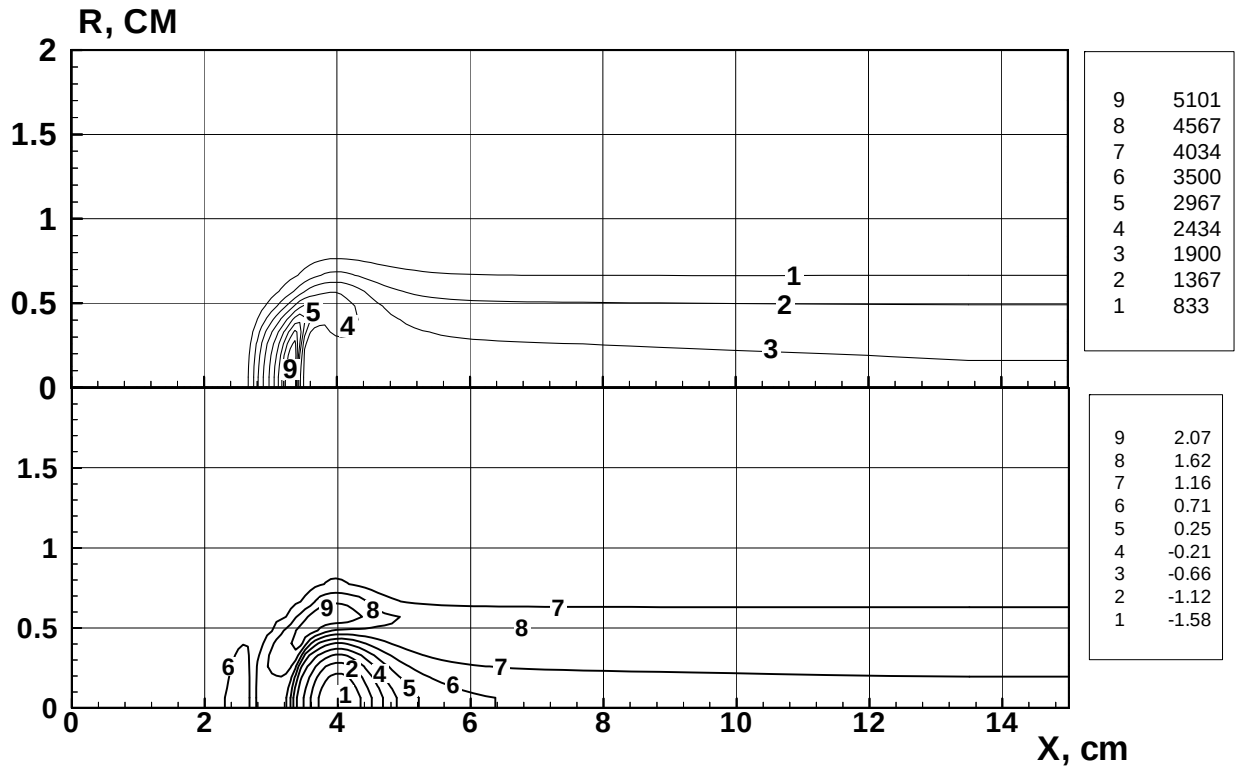

**Fig.5.** Temperature and axial velocity distributions at $u_0 = 40$ m/s and $Q_V = 3.0$ kW/cm$^3$. Set of calculations with **increasing** heat release

Figure 6 shows temperature and velocity distribution at $u_0 = 40$ m/sec and $Q_V = 21.0$ kW/cm$^3$. In this case the maximal temperature amounts to 23000 K.

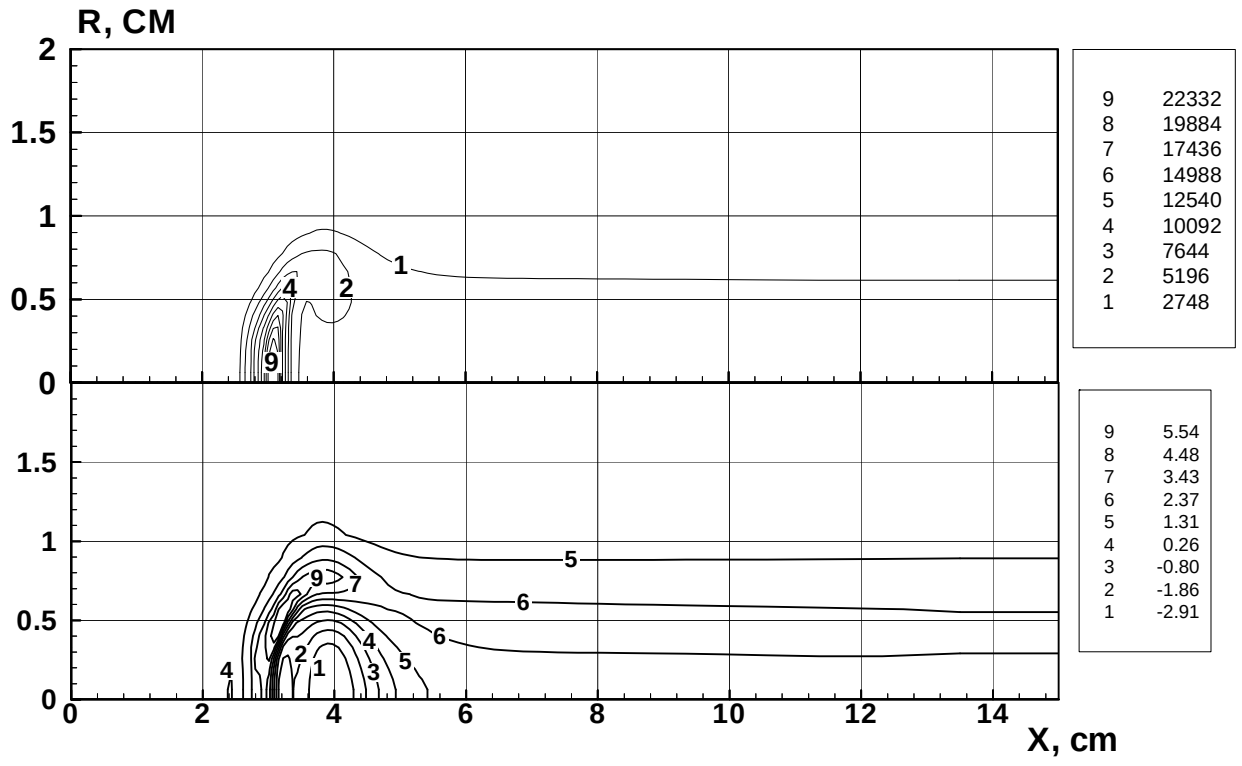

**Fig.6.** Temperature and axial velocity distributions at $u_0 = 40$ m/s and $Q_V = 21.0$ kW/cm$^3$. Set of calculations with **increasing** heat release

Numerical simulation results shown in Figs. 7 and 8 were obtained for the case with gradual reduction of $Q_V$ with the step $\Delta Q_V = -0.5$ kW/cm$^3$, when the vortical solution was taken as the initial one. Figure 7 shows temperature and velocity distributions at $Q_V = 2.25$ kW/cm$^3$, and Fig. 8 shows the data at $Q_V = 2.0$ kW/cm$^3$. Note that in the set of calculations with increasing $Q_V$ the laminar solution was obtained at $Q_V = 2.5$ kW/cm$^3$. Thus at the considered conditions it is possible to specify the range of values of $Q_V$ inside which there is the bifurcation of the solution: $u_0 = 40$ m/sec, $Q_V^{**} = 2.25 < Q_V < Q_V^* = 2.5$ kW/cm$^3$. At other speeds these bifurcational ranges are found to be:

a) $u_0 = 50$ m/sec,

$Q_V^{**} = 2 < Q_V < Q_V^* = 3$ kW/cm$^3$;

b) $u_0 = 100$ m/sec,

$Q_V^{**} = 3.5 < Q_V < Q_V^* = 5.5$ kW/cm$^3$;

c) $u_0 = 150$ m/sec,

$Q_V^{**} = 5.5 < Q_V < Q_V^* = 7$ kW/cm$^3$,

d) $u_0 = 200$ m/sec,

$Q_V^{**} = 7.5 < Q_V < Q_V^* = 9$ kW/cm$^3$.

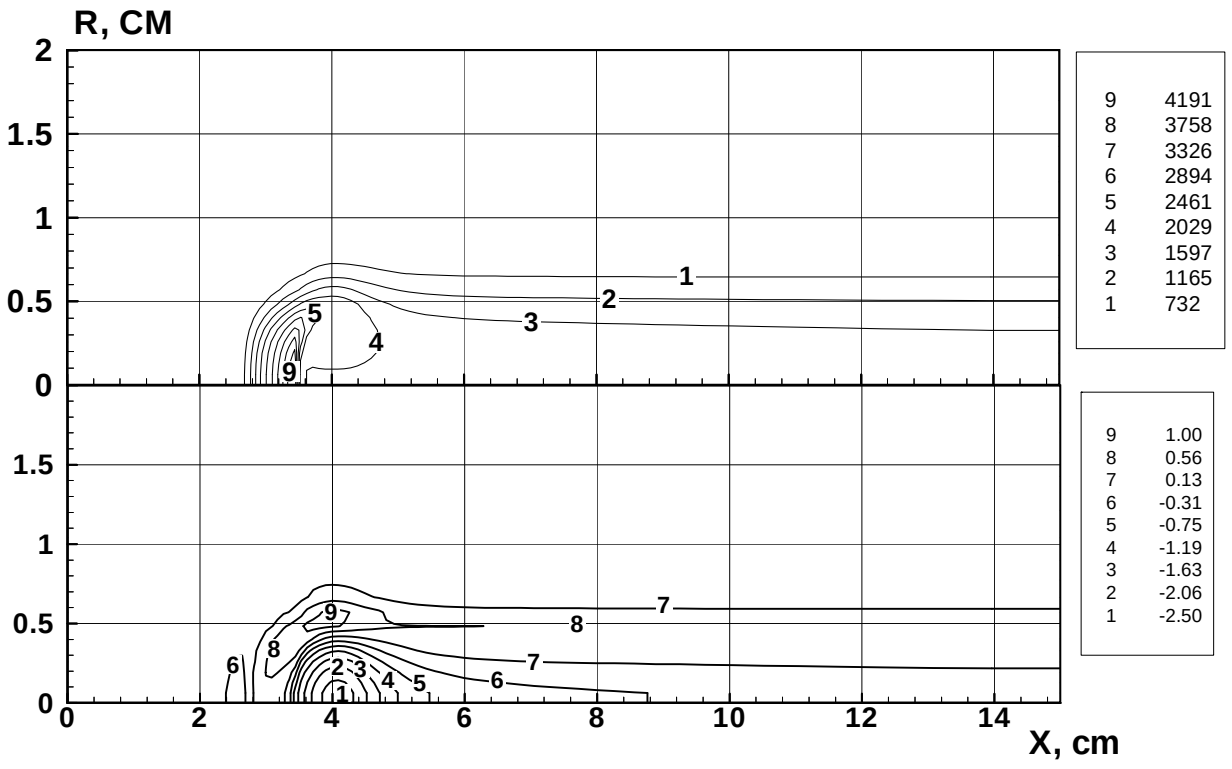

**Fig.7.** Temperature and axial velocity distributions at $u_0 = 40$ m/s and $Q_V = 2.25$ kW/cm$^3$. Set of calculations with **decreasing** heat release

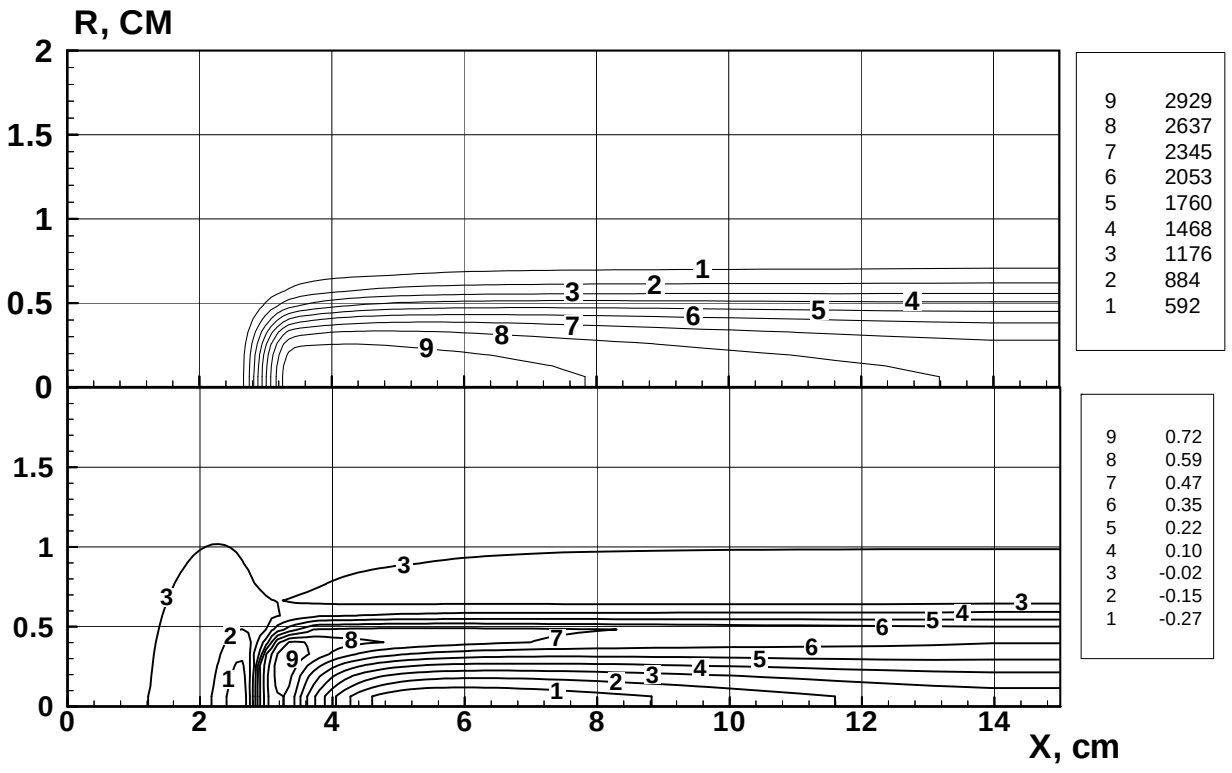

**Fig.8.** Temperature and axial velocity distributions at $u_0 = 40$ m/s and $Q_V = 2.0$ kW/cm$^3$. Set of calculations with **decreasing** heat release

Numerical simulation results for the set of calculations with $u_0 = 30$ m/sec are particular of interest for analysis of the unsteady gas dynamic structures. Remember that at smaller speeds the phenomena of a flow bifurcation was not revealed. The specified speed $u_0 = 30$ m/sec is near the bottom border of the range of speeds, within which the bifurcation was found. This case is characterized by significant instability in the calculated data. Therefore the calculations at this specified speed were performed with different values of $\Delta Q_V$.

In the first set of calculations the step of increasing of the heat release power was $\Delta Q_V = 0.5$ kW/cm$^3$, in the second set – $\Delta Q_V^{\mathrm{II}} = 0.1$ kW/cm$^3$, and in the third set $-\Delta Q_V^{\mathrm{III}} = 0.025$ kW/cm$^3$. The transition from laminar to vortical motion was observed in the range $Q_V = 1.5 \div 2$ kW/cm$^3$ in the first case, and in the range – $Q_V = 3.7 \div 3.8\,\text{kW/cm}^3$ in the second case. In the third case the transition from laminar to vortical movement was found in the range $Q_V = 5.925 \div$ $\div 5.95\,\text{kW/cm}^3$. Figure 9 shows temperature and velocity distributions for laminar mode at $u_0 = 30$ m/sec and $Q_V = 3.7\,\text{kW/cm}^3$, and Fig. 10 shows the same data for vortical mode at $u_0 = 30$ m/sec and $Q_V = 3.8\,\text{kW/cm}^3$).

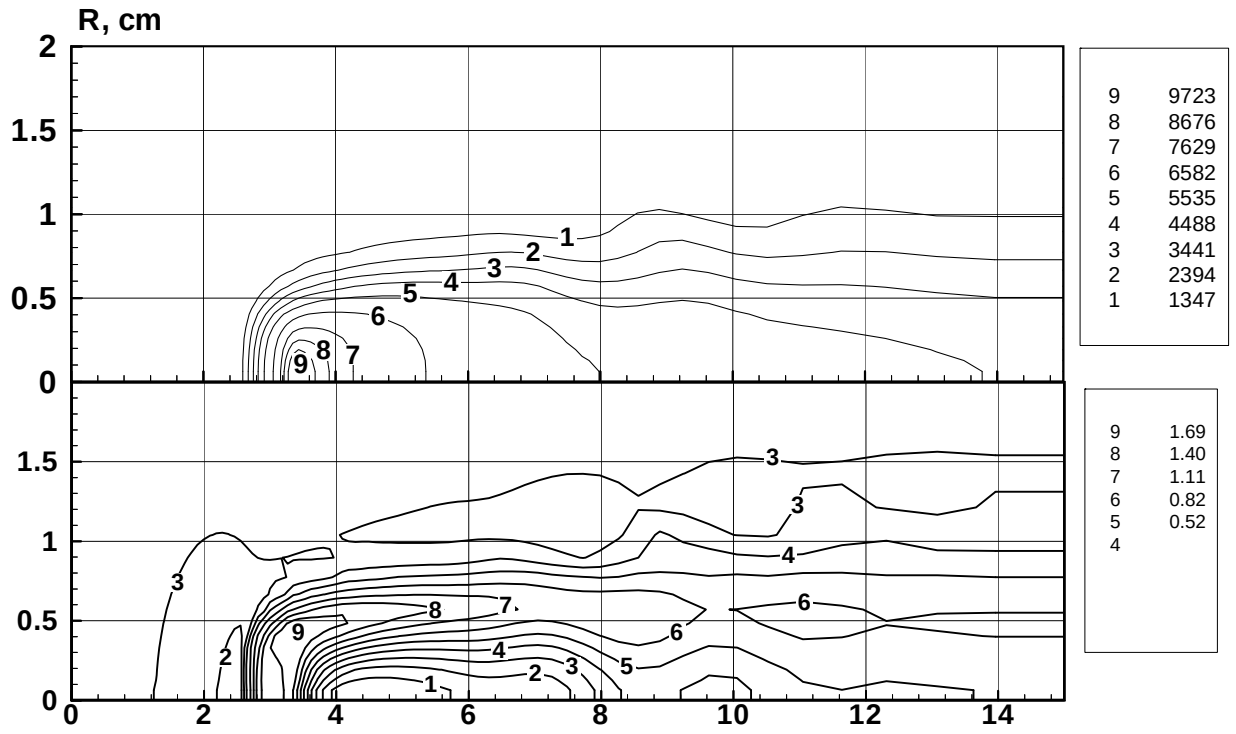

**Fig.9.** Temperature and axial velocity distributions at $u_0 = 30$ m/s and $Q_V = 5.925\,\text{kW/cm}^3$. Set of calculations with **increasing** heat release ( $\Delta Q_V = +0.025\,\text{kW/cm}^3$)

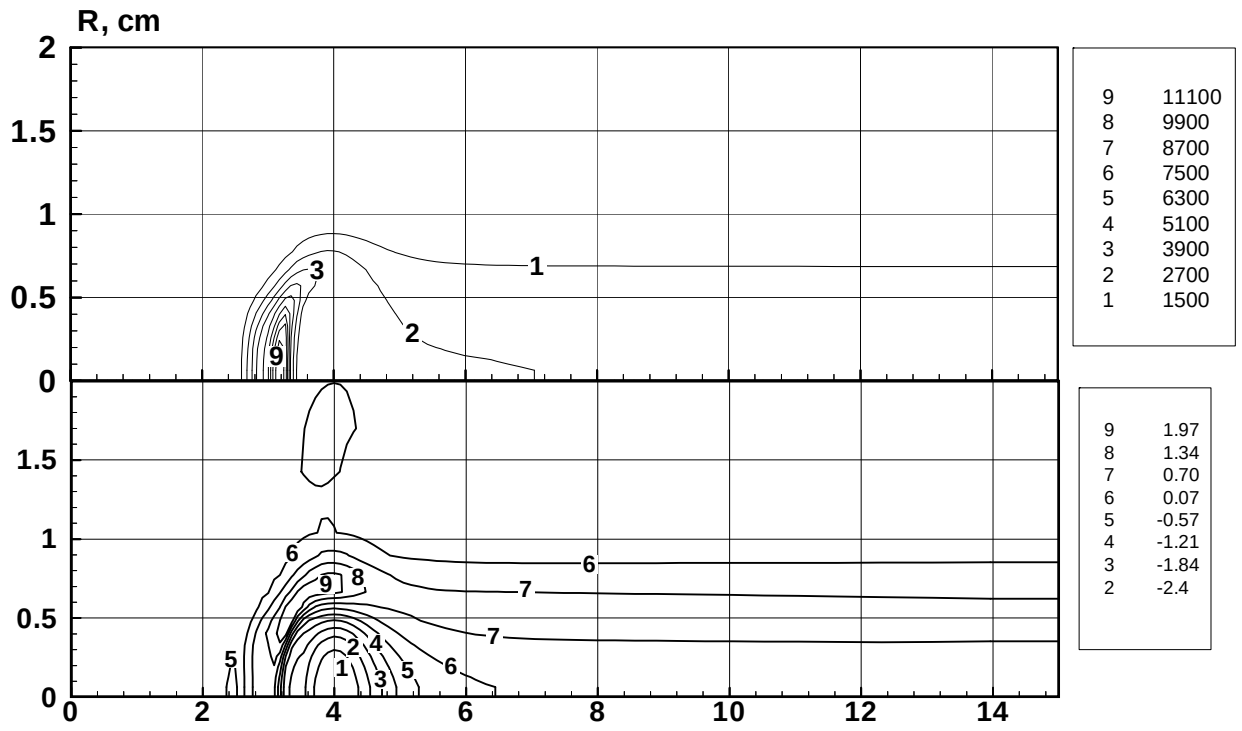

**Fig.10.** Temperature and axial velocity distributions at $u_0 = 30$ m/s and $Q_V = 5.95\,\text{kW/cm}^3$. Set of calculations with **increasing** heat release ( $\Delta Q_V = +0.025\,\text{kW/cm}^3$)

Figure 11 shows the axial velocity distribution for laminar and vortical solutions obtained with increasing heat release power for this case. It indicates that the reducing level of heat release disturbance results in essentially larger values of the heat release capacity at the transition from laminar to vortical flow. It is reasonable that the gas is heated up to the large temperatures. For example, at the greatest heat release capacity $Q_V = 5.925\,\text{kW/cm}^3$, at which the laminar solution was observed the temperature inside heated area achieves $T = 10000$ K. The distributions of temperature and axial velocity $u$ for this case are shown in Fig. 9. Appropriate distributions of the functions after transition to vortical motion ( $Q_V = 5.95\,\text{kW/cm}^3$) are shown in Fig. 10. Maximal temperature in this case reaches 12000 K.

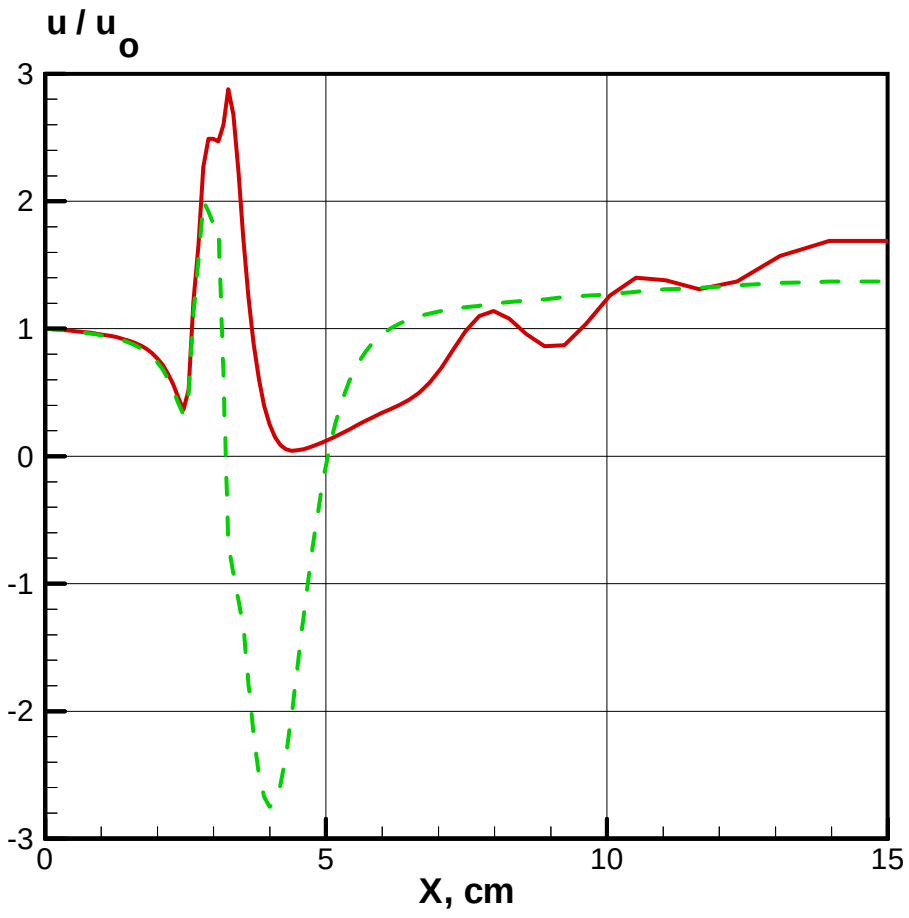

**Fig.11.** Axial velocity distributions at $r = 0$; $u_0 = 30$ m/s and $Q_V = 5.925$ (solid line) and 5.95 (dashed line) $\text{kW/cm}^3$. Set of calculations with **increasing** heat release

The discussed case of calculations is also remarkable since the steady-state laminar solution does not exist at $Q_V = 5.925\,\text{kW/cm}^3$. Self-oscillatory process with periodic variations of velocity components and temperature is observed in this case. Partly it can be seen from distributions of temperature and speed (see Fig. 9). Actually an instant photo is shown here.

## 5 Conclusion

Numerical investigation of subsonic flows with localized heat release regions showed that at certain conditions in gas flows it is possible to detect a bifurcation of gas dynamic structure. Namely, two different quasi-steady-state gas dynamic configurations at the same initial data do exist.